\documentclass[12pt]{article}

\usepackage{euscript,amsfonts,amssymb,epsfig,latexsym}

\def\<{\langle}
\def\>{\rangle}

\def\Kc{{\cal K}}
\def\Pc{{\cal P}}

\def\pd{\partial}

\title{Numerical Study of Nonlinear Equations with Infinite Number of Derivatives}

\author{
Yaroslav Volovich\\
Physics Department, Moscow State University\\
Vorobievi Gori, 119899 Moscow, Russia\\
email: yaroslav@aylabs.com
}
\date{}

\begin{document}

\maketitle

\begin{abstract}
We study equations with infinitely many derivatives.
Equations of this type form a new class of equations in mathematical
physics. These equations originally appeared in p-adic and later in
fermionic string theories and their investigation is of much interest
in mathematical physics and applications, in particular in cosmology.
Differential equation with infinite number of derivatives could
be written as nonlinear integral equations. We perform numerical
investigation of solutions of the equations.
It is established that these equations have two different regimes
of the solutions: interpolating and periodic.
The critical value of the parameter $q$ separating these regimes
is found to be $q^2_{cr}\approx 1.37$.
Convergence of iterative procedure for these equations is proved.
\end{abstract}

\newpage
\tableofcontents

\newpage
\section{Introduction}

Normally, in mathematical physics one considers equation with finite
number of derivatives. For such equations there are well developed methods
of solution for various boundary problems, see for example \cite{Vla}.
Recently in works on p-adic and then in real string theories a certain
class of nonlinear equations which involve infinite number of derivatives
is started to be explored \cite{BFOW}-\cite{Sen}. Such equations form
an important new class of equations in mathematical physics. New methods
to study uniqueness and existence  of solution should be developed. It
is not clear apriori how to pose boundary or initial value problems for
these equations.

An example of new equations has the form
\begin{equation}
\label{infdiff}
e^{a\Delta}\varphi=\varphi^k,
\end{equation}
where $\Delta$ is the Laplace (or D'Alamber) operator, $a$ is a real parameter,
and $k$ is nonnegative integer. This equation was originally studied in p-adic string
theory. Soliton solutions to (\ref{infdiff}) where considered in \cite{BFOW,FramPadic}.
It is interesting to study equation (\ref{infdiff}) in the simplest case when
$\varphi=\varphi(t)$ depends only on one real variable $t$ \cite{MolZw,Sen}.
In this case we can rewrite (\ref{infdiff}) in the form of integral equation
\begin{equation}
\label{int-1}
(\Kc\varphi)(t)=\varphi(t)^k,
\end{equation}
where
\begin{equation}
\label{intdiff}
e^{a\pd_t^2}\varphi(t)=
(\Kc\varphi)(t)=
  \frac{1}{\sqrt{4a\pi}}\int_{-\infty}^{\infty}e^{-\frac{(t-t')^2}{4a}}\varphi(t') dt'
\end{equation}
In recent paper \cite{MolZw} a kink solution found in \cite{BFOW} was confirmed
and also oscillatory solutions where found.

In this paper the equations of the form (\ref{int-1}) and more general
equations are investigated.
These equations describe dynamics of the scalar field with the lowest
mass square (tachyon field) in fermionic string model \cite{AJK}.
We perform numerical investigation of solutions of these equations.
It is established that these equations have two different regimes
of the solutions: interpolating and periodic.
The critical value of the parameter $q$ separating these regimes
is found to be $q^2_{cr}\approx 1.37$.
Convergence of iterative procedure for these equations is proved.
To construct numerical algorithm we essentially used object-oriented
design, please see section \ref{num1} for further details.

This paper is organized as follows. In the next section we study the
equation taken from p-adic string model, we prove the convergence of
iterative procedure for this equation.
In the next two sections we study equations describing the scalar
field (tachyon field) in fermionic string.
Equations for fermionic string generalize the equations for p-adic
string.
Finally, in the last section we provide physical details including
actions for which corresponding equations of motion form the
subject of this paper.

\section{Scalar Field Dynamics in P-adic String Model}

In this section we study the following nonlinear integral
equation
\begin{equation}
\label{p-adic}
(\Kc\varphi)(t)=\varphi(t)^p,
\end{equation}
where integral operator $\Kc$ is defined by
\begin{equation}
\label{kc-def}
(\Kc\varphi)(t)=\frac{1}{\sqrt\pi}\int_{-\infty}^{\infty} e^{-(t-t')^2}\varphi(t') dt'
\end{equation}
Originally $p$ in the rhs of (\ref{p-adic}) is a prime number, although
here it is not important.
This equation has several physical applications basically it describes
dynamics of homogenous scalar field configurations in the p-adic string model
\cite{BFOW,FramPadic,VVZ}
We describe the physics behind in more detail in the section \ref{phys}.
The equation (\ref{p-adic}) for $p=3$ was studied in \cite{BFOW}.
It was numerically shown that it has solution which goes to $\mp 1$ on
infinities. In this section we prove the convergence of iterative procedure
used in \cite{BFOW} to construct numerical solution of (\ref{p-adic}).
In \cite{MolZw} it was shown that (\ref{p-adic}) does not have
monotonic solutions for even $p$. In the text below we discuss only the case
$p=3$ since it is the most illustrative for the p-adic string model and provides
an approximation for fermionic string model.

We consider equation (\ref{p-adic}) for the case $p=3$
\begin{equation}
\label{p-adic2}
(\Kc\varphi)(t)=\varphi(t)^3
\end{equation}
We are searching for the solution which has constant
asymptotic behavior on infinities. For the case of constant
field equation (\ref{p-adic2}) takes the form
\begin{equation}
\label{vac-eq}
\varphi_0=\varphi_0^3
\end{equation}
It has three solutions
\begin{equation}
\label{vacuums}
\varphi_0^{(1)}=1,~~~~\varphi_0^{(2)}=0,~~~~\varphi_0^{(3)}=-1
\end{equation}
We are interested in odd solution of (\ref{p-adic2}) which goes to
$\mp 1$ on infinities, i.e.
\begin{equation}
\label{asympt}
\lim_{t\to\pm\infty}\varphi(t)=\mp 1
\end{equation}
and
\begin{equation}
\label{parity}
\varphi(t)=-\varphi(-t)
\end{equation}

\subsection{Construction of Solution Using Iterative Procedure}
\label{iter-p}

To construct solution of integral equation (\ref{p-adic2}) with
the properties (\ref{asympt})-(\ref{parity}) one could use
the following iterative procedure \cite{BFOW,Kr}
\begin{equation}
\label{iter}
\varphi_{n+1}=(\Kc\varphi_n)^{1/3},
\end{equation}
where zero approximation is taken as
\begin{equation}
\label{iter-0}
\varphi_{0}(t)=-\varepsilon(t),
\end{equation}
where $\varepsilon(t)$ is a step function defined by
\begin{equation}
\label{step}
\varepsilon(t)=\left\{
  \begin{array}{l}
   ~-1,~~\mbox{for}~~t<0\\
   ~~~~0,~~\mbox{when}~~t=0\\
   ~~~~1,~~\mbox{for}~~t>0
  \end{array}
\right.
\end{equation}
Note that in the rhs of (\ref{iter-0}) expression of the
forme $a^{1/3}$ denotes an arithmetic cubic root of $a$
which is well defined for negative arguments as
\begin{equation}
a^{1/3}=\left\{
  \begin{array}{l}
   a^{1/3},~~~~~~a\geq 0\\
   -|a|^{1/3},~~a<0
  \end{array}
\right.
\end{equation}

\begin{figure}[!ht]
\centering
\includegraphics[width=7cm]{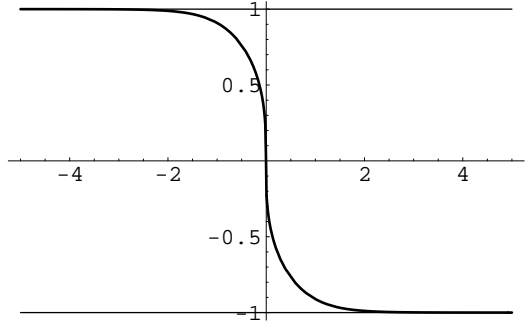}
\caption{Results of iterative procedure (\ref{iter})-(\ref{iter-0}) for a large
number of steps ($\sim 10^5$).}
\label{fig:padic}
\end{figure}

Results of the iterative procedure (\ref{iter})-(\ref{iter-0})
are presented on fig.\ref{fig:padic}.

In the next section we will prove convergence of iterative procedure
(\ref{iter})-(\ref{iter-0}) that supports the results of numerical
computations in \cite{BFOW}.

The first approximation $\varphi_{1}(t)$ could be computed analytically
and is given by the arithmetical cubic root of the error function.
Indeed,
\begin{equation}
\label{iter-1}
\varphi^3_{1}(t)=(\Kc\varphi_0)(t)=-\mbox{erf}(t),
\end{equation}
where
$$
\mbox{erf}(x) = \frac{2}{\sqrt\pi}\int_0^x \exp(-t^2)dt
$$

\begin{figure}
\centering
\includegraphics[width=4cm]{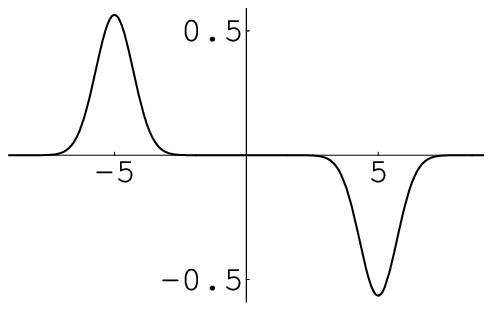}
\includegraphics[width=4cm]{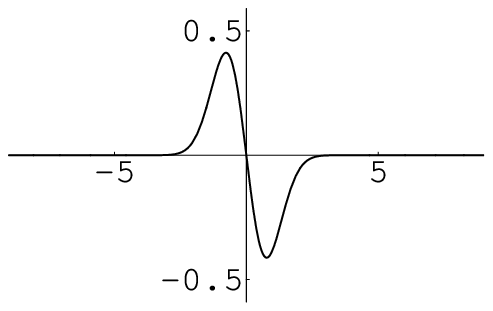}
\includegraphics[width=4cm]{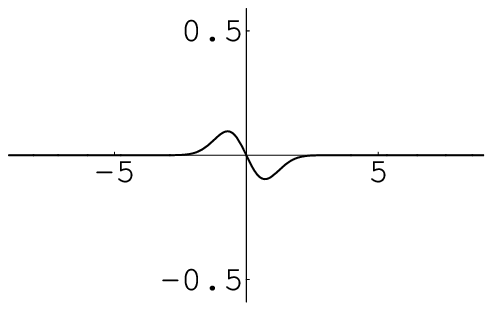}
\caption{The kernel $K_-(b,t)$ as a function of $t$ for $b=-5,-0.5,-0.1$.
It is positive for $t<0$, $b<0$.}
\label{fig:k}
\end{figure}

In order to prove the convergence of iterative procedure
(\ref{iter}) it is convenient to take into account the parity
property (\ref{parity}). This allows us to rewrite equation
(\ref{p-adic2}) on the semi-axis
\begin{equation}
\label{half-axis}
(\Kc_{-}\varphi)(t)=\varphi(t)^3,~~t<0
\end{equation}
where integral operator $\Kc_{-}$ is defined by
\begin{equation}
\label{Kc-def}
(\Kc_{-}\varphi)(t)=\int_{-\infty}^{0} K_-(t,t')\varphi(t')dt',
\end{equation}
where the kernel $K_{-}(t,t')$ is given by
\begin{equation}
\label{K-minus-def}
K_{-}(t,t')=\frac{1}{\sqrt{\pi}}\left[e^{-(t-t')^2}-e^{-(t+t')^2}\right]
\end{equation}
Let us note that there is the following positiveness property of the kernel
\begin{equation}
\label{Kpos}
K_-(t,t')>0 ~~\mbox{for all} ~~ t<0,~t'<0
\end{equation}
and
$$
K_-(t,0)=K_-(0,t')=0,
$$
see fig.\ref{fig:k}.

\subsection{Convergence of Iterative Procedure}
\label{conv}

Here we describe the construction of solution
for equation (\ref{half-axis}) on the semi-axis ($t<0$).
Consider the following iterative procedure
\begin{equation}
\label{iter-half}
\varphi_{n+1}=\left(\Kc_{-}\varphi_n\right)^{1/3}
\end{equation}
Let us note, that the second iteration
\begin{equation}
\label{2iter}
\varphi_{2}=\left(\Kc_{-}\varphi_1\right)^{1/3}
\end{equation}
could be represented in the form
\begin{equation}
\label{2iter-rep}
\varphi_{2}(t)=\varphi_{1}(t)(1-\Delta(t)),
\end{equation}
where $\Delta(t)$ is defined by
\begin{equation}
\label{delta}
\Delta(t)=\frac{\varphi_{1}(t)-\varphi_{2}(t)}{\varphi_1(t)}
\end{equation}
\begin{figure}[!ht]
\centering
\includegraphics[width=5.9cm]{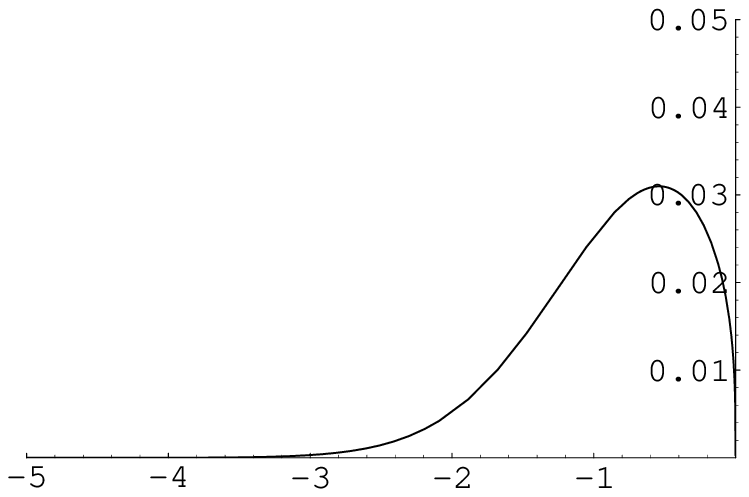}
\includegraphics[width=5.9cm]{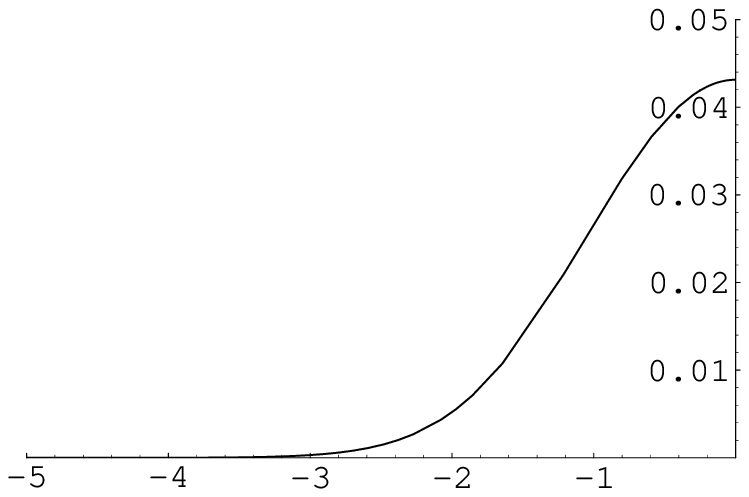}
\caption{Plots of $\varphi_{1}(t)-\varphi_{2}(t)$ and $\Delta(t)$ illustrate that
$\Delta(t)<0.05$.}
\label{fig:delta}
\end{figure}
On fig.\ref{fig:delta} the difference $(\varphi_{1}(t)-\varphi_{2}(t))$
and $\Delta(t)$ are shown for negative $t$. From
(\ref{iter-1}), (\ref{2iter}), and (\ref{delta}) it follows
that (see. fig.\ref{fig:delta})
\begin{equation}
\label{delta-lim}
\Delta(t)<\Delta_{max}=0.05
\end{equation}
thus
\begin{equation}
\label{equal}
\varphi_1(x)>\varphi_2(x)>\varphi_1(x)(1-\Delta_{max})
\end{equation}
Since we have the positiveness property (\ref{Kpos})
we can integrate the inequality (\ref{equal}) with the kernel $K_{-}(t,t')$
\begin{equation}
\label{equal-int}
\int _{-\infty}^{0} K_{-}(y,x)\varphi_1(x)dx \geq\int
_{-\infty}^{0} K_-(y,x)\varphi_2(x)dx \geq
\end{equation}
$$
\geq(1-\Delta_{max})\int _{-\infty}^{0} K_-(y,x)\varphi_1(x)dx
$$
Inequality (\ref{equal-int}) leads us to
\begin{equation}
\label{equal-3}
\varphi_2^3(y)\geq \varphi_3^3(y)\geq\varphi_2^3(y)(1-\Delta_{max})
\end{equation}
Now taking the arithmetical cubic root we get
\begin{equation}
\label{equal-3sq}
\varphi_2(y)\geq \varphi_3(y)\geq\varphi_2(y)(1-\Delta_{max})^{1/3}
\end{equation}
and more over
\begin{equation}
\label{equal-13sq}
\varphi_1(y)\geq \varphi_2(y)\geq \varphi_3(y)\geq
\varphi_2(y)(1-\Delta_{max})^{1/3}\geq \varphi_1(y)(1-\Delta_{max})^{1+1/3}
\end{equation}
Analogously we have
\begin{equation}
\label{equal-n} \varphi_1(y)\geq \varphi_n(y)\geq
\varphi_1(y)(1-\Delta_{max})^{1+1/3+(1/3)^2+...+(1/3)^{n-2}}=
\end{equation}
$$
=\varphi_1(y)(1-\Delta_{max})^{\frac{3}{2}-\frac{1}{2}(\frac{1}{3})^{n-2}},
$$
i.e.
\begin{equation}
\label{ineq-n1}
\varphi_1(y)\geq
\varphi_n(y)\geq
\varphi_1(y)(1-\Delta_{max})^{\frac{3}{2}-\frac{1}{2}(\frac{1}{3})^{n-2}},
\end{equation}
From (\ref{ineq-n1}) it follows that $\varphi_n(y)$ is uniformly bounded
on the whole negative semi-axis.

More over, beside (\ref{equal-3}) and (\ref{equal-3sq}) we have
\begin{eqnarray}
\label{equal-n1-s}
(1-\Delta_{max})^{1/3}\varphi^3_3(y)\leq\varphi^3_4(y)\leq\varphi^3_3(y)
\end{eqnarray}
thus
\begin{eqnarray}
\label{equal-n1-s-2}
(1-\Delta_{max})^{1/9}\varphi_3(y)\leq\varphi_4(y)\leq\varphi_3(y)
\end{eqnarray}
Analogously we have
\begin{eqnarray}
\label{equal-n1-ss-2}
(1-\Delta_{max})^{1/3^{n-1}}\varphi_n(y)\leq\varphi_{n+1}(y)\leq\varphi_n(y)
\end{eqnarray}
Hence
\begin{eqnarray}
\label{equal-n1-ss}
|\varphi_n(y)-\varphi_{n+1}(y)|\leq\varphi_n(y)\left(1-(1-\Delta_{max})^{\frac{1}{3^{n-1}}}\right)
\end{eqnarray}
Finally, taking into account that $\varphi_n(y)$ is uniformly bounded
we obtain
\begin{eqnarray}
\label{req}
|\varphi_n(y)-\varphi_{n+1}(y)|<\frac{C}{3^n},
\end{eqnarray}
where $C$ is constant.

From (\ref{req}) it follows uniform convergence of $\varphi_n(y)$ on the semi-axis,
that is uniform convergence of the iterative procedure on the semi-axis.

\section{Scalar Field Dynamics in Fermionic String Model I}
\label{fermI}

\subsection{Integro-Differential Equation}

In this section we consider the following integro-differential equation
\begin{equation}
\label{ferm}
(-q^2\pd_t^2+1)(\Kc\varphi)(t)=\varphi(t)^3,
\end{equation}
where the integral operator $\Kc$ is defined by (\ref{kc-def})
and $q$ is a parameter.
This equation describes dynamics of the homogenous scalar field
with the lowest mass square (tachyon) in fermionic string model
in some approximation, please see section \ref{phys} for more
physical details. The equation (\ref{ferm}) transforms to
p-adic equation (\ref{p-adic2}) in the case $q=0$.
Although the value of parameter $q$ for fermionic string model
is given by
\begin{equation}
q^2_{string}=-\frac{1}{4\ln\frac{4}{3\sqrt3}}\approx 0.96,
\end{equation}
we consider equation (\ref{ferm}) for various values of parameter $q$.

\subsection{Fully Integral Form and Iterative Procedure}

The equation (\ref{ferm}) could be written in fully integral form
\begin{equation}
\label{ferm-integ}
(\Kc_q\varphi)(t)=\varphi(t)^3,
\end{equation}
where the kernel of integral operator $\Kc_q$ is obtained
by differentiating the kernel of (\ref{kc-def}), namely
by applying operator $(-q^2\pd_t^2+1)$ to $\exp[-(t-t')^2]$
\begin{equation}
\label{integ-q}
(\Kc_q\varphi)(t)=\frac{1}{\sqrt{\pi}}
  \int_{-\infty}^{\infty}
    e^{-(t-t')^2}[1+2q^2(1-2(t-t')^2)]\varphi(t')dt'
\end{equation}
As in the previous section we are interested in odd
solution of (\ref{ferm-integ}) which goes to $\mp 1$ on infinities,
i.e. the solution of (\ref{ferm-integ}) with the properties
(\ref{asympt})-(\ref{parity}).

To construct the solution we use the following iterative procedure
\begin{equation}
\label{ferm-iter}
\varphi_{n+1}=(\Kc_q\varphi_n)^{1/3},
\end{equation}
where zero approximation is taken as
\begin{equation}
\label{ferm-iter-0}
\varphi_{0}(t)=-\varepsilon(t),
\end{equation}
and $\varepsilon(t)$ is a step function defined by (\ref{step}).
In the previous section we proved that this iterative procedure
converges to a solution for the case $q=0$.

\subsection{Numerical Results}
\label{num1}

We use the iterative procedure (\ref{ferm-iter})-(\ref{ferm-iter-0})
to perform numerical investigation of the solution
properties for various values of parameter $q$.

\begin{figure}
\centering
\includegraphics[width=7cm]{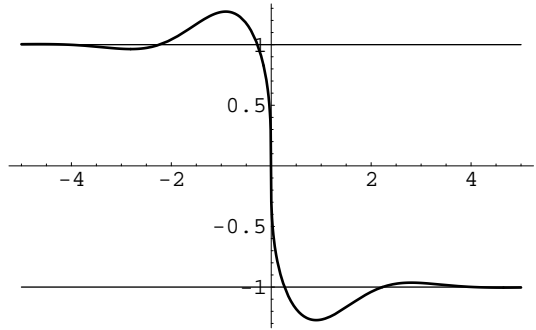}
\caption{Results of iterative procedure (\ref{ferm-iter})-(\ref{ferm-iter-0})
for $q=q_{string}$.}
\label{fig:q2=0.96}
\end{figure}

It was found that there exists a critical value of the parameter
$q^2$: $q^2_{cr}\approx 1.37$. This is the maximum value of $q$
for which there exist interpolating solutions. For $q^2<q^2_{cr}$
there were numerically found solutions with asymptotic behavior
(\ref{asympt}) which oscillate along $-\varepsilon(t)$ with
exponentially decreasing amplitude. On fig.\ref{fig:q2=0.96}
it is demonstrated a numerical limit of iterative procedure
(\ref{ferm-iter})-(\ref{ferm-iter-0}) for $q=q_{string}$.

\begin{figure}
\centering
\includegraphics[width=10cm]{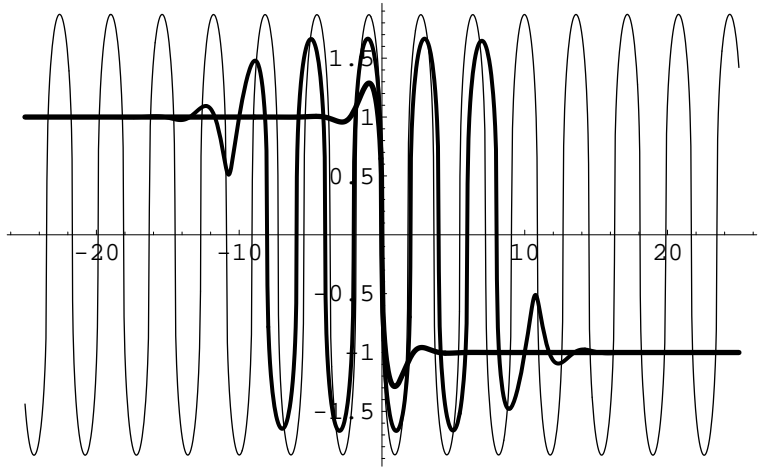}
\caption{Appearance of the swing of the solution with the increase of $q$.
The case $q^2=1$ (thick line) -- no swing, interpolating regime,
$q^2=1.4$ (medium-thickness line) -- {\it appearance} of the swing,
oscillatory regime, $q^2=1.8$ (thin line) -- oscillatory regime.}
\label{fig:osc}
\end{figure}

\begin{figure}
\centering
\includegraphics[width=10cm]{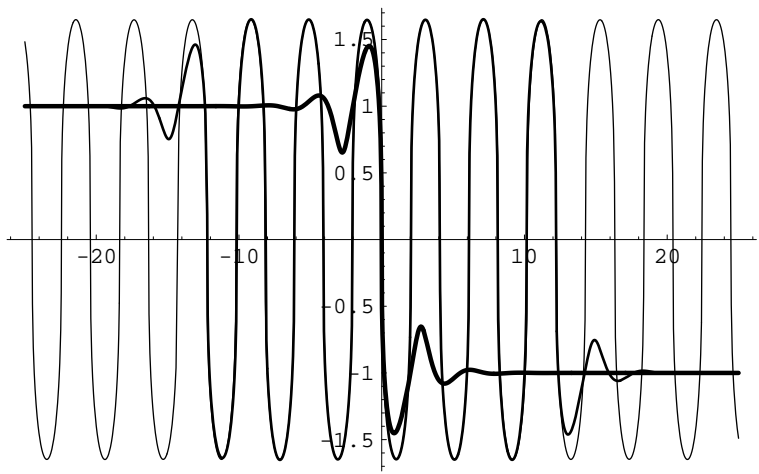}
\caption{Transformation to the periodic regime for $q^2=1.38$ appears
on quite big step numbers -- $100$-th step (thick line), $250$-th step
(medium thickness line), $500$-th step (thin line).}
\label{fig:large-iter}
\end{figure}

While increasing the value of $q$ above the critical value $q_{cr}$
there appears a swing and the solution of (\ref{ferm-integ})
becomes periodic (see fig.\ref{fig:osc}).
Let us note that this swing for $q^2\simeq 1.37\pm 0.01$ appears on
quite big step numbers (values of $n$ in (\ref{ferm-iter})).
On fig.\ref{fig:large-iter} the result of iterative procedure
for $q^2=1.38$ for various step numbers are shown.

Numerical investigation of iterative procedure on very high step numbers
($\sim 10^5$) showed that transformation to periodic regime does not
appear for $q<q_{cr}$. Although for values of $q$ a little above
the critical value $q_{cr}$ the swing, i.e. a transformation to
periodic regime appears on much smaller step numbers ($\sim 10^2$).
This shows that there is a principle difference between solutions
for $q$ below and above $q_{cr}$. This fact is discussed in more
details in the next section.

Here we used (\ref{ferm-iter-0}) as a zero approximation $\varphi_0$
which is not continuous at the point $t=0$. To understand that this
discontinuity does not affect the results of iterative procedure we
tried several smooth continuous zero approximations which
have asymptotic behavior (\ref{asympt}). In particular we tried
$-\frac{2}{\pi}\arctan(t)$ and $-\mbox{erf}(t)$.
The results of the iterative procedure with these zero
approximations where for large step numbers the same
as with (\ref{ferm-iter-0}).

For numerical computation we used the following approximation for
$\Kc$
\begin{equation}
\label{Kcapp}
(\Kc\varphi)(t)\cong
  \frac{1}{\sqrt{\pi}}\int\limits_{t-\Delta}^{t+\Delta}
    e^{-(t-t')^2}\varphi(t') dt'
\end{equation}
Here $\Delta$ was automatically adjusted, namely $\Delta\simeq 10$.
This approximation lead us to the algorithm with linear complexity
that allowed us to compute $\sim 10^5$ iterations on parallel 64-bit
parallel Sun Ultra-SPARC machine. The entire algorithm was written
in C++.

To construct numerical algorithm we essentially used object-oriented
design \cite{Gamma,Eliens}. This allowed us to develop a general
algorithm which solves some class of integral equations, in
particular it was used to solve a system of nonlinear integral
equations discussed in section \ref{fermII}.

\newpage
\subsection{Two Regimes of The Solution}
\label{2reg}

In this section we are interested in the mechanism which forms
numerically found exponentially decreasing oscillations and the
presence of $q_{cr}$. The basic idea is to present a solution with
nonzero $q$ as a deviation along the solution with $q=0$.
We write linear equation for this deviation for large values of $t$.

Let us write the solution of (\ref{ferm}) as a sum
\begin{equation}
\label{excit}
\varphi(t)=\phi_0(t)+\chi(t),
\end{equation}
where $\phi_0(t)$ denotes the solution of (\ref{p-adic}).
Substituting (\ref{excit}) to (\ref{ferm}) and leaving
only linear terms in $\chi$ we get
\begin{equation}
\label{chi1}
(-q^2\pd_t^2+1)\Kc(\phi_0+\chi)=\phi_0^3+3\phi_0^2\chi
\end{equation}
Now using the fact that $\phi_0$ satisfies (\ref{p-adic}) we get
the linear integro-differential equation on $\chi(t)$
\begin{equation}
\label{chi-lin}
(-q^2\pd_t^2+1)\Kc\chi=3\phi_0^2\chi+q^2\pd_t^2\Kc\phi_0
\end{equation}
From (\ref{Kcapp}) we see that in $\Kc$ integration could
be taken in finite limits with good precision. This gives
us ability to write large $t$ approximation of (\ref{chi-lin})
\begin{equation}
\label{chi-smp}
(-q^2\pd_t^2+1)\Kc\chi=3\chi,
\end{equation}
here we used the fact that $\phi_0(t)\simeq 1$ and its derivatives
are zero for large $t$. Representing $\chi(t)$ as
$$
\chi(t)=e^{i\Omega t}
$$
and substituting it to (\ref{chi-smp}) we get the following
characteristic equation
\begin{equation}
\label{Omega-q}
(q^2\Omega^2+1)e^{-\frac{1}{4}\Omega^2}=3
\end{equation}
We consider (\ref{Omega-q}) as an equation for complex variable
$\Omega$ with parameter $q^2$.
There is a minimum value of $q^2$: $q^2_0\approx 1.77$ for which
$\Omega$ is real.
For $q^2<q^2_0$ it has solutions
with nonzero imaginary parts which gives oscillatory regime with
exponentially decreasing amplitude. For $q^2>q^2_0$ solutions are
real. As it was mentioned in the previous section numerical
computations give the critical value $q^2_{cr}\approx 1.37$
which is smaller than found above value $q^2_0\approx 1.77$.
This probably means that one can not neglect the difference
of $\phi_0$ from $1$ in the rhs of (\ref{chi-lin}) or even
one has to take into account nonlinear terms in $\chi$ in (\ref{chi1}).
Nevertheless method discussed in this section provides a good
qualitative explanation to the behavior of solutions for
small values of $q$.

\subsection{Periodic Solutions For Large $q$}
\label{big-q}

In this section we investigate the asymptotic behavior of the solution
for large $q$. First, let us introduce a new function $\chi(t)$ defined by
\begin{equation}
\label{chi-Psi}
\chi(t)=\varphi(qt),
\end{equation}
In terms of $\chi(t)$ the equation (\ref{ferm}) rewrites as
\begin{equation}
\label{chi-int}
(-\pd_t^2+1)(\Pc_q\chi)(t)=\chi(t)^3,
\end{equation}
where integral operator $\Pc_q$ is defined by
\begin{equation}
\label{Pc}
(\Pc_q\chi)(t)=\frac{q}{\sqrt{\pi}}
  \int_{-\infty}^\infty
  e^{-q^2(t-t')^2}
  \chi(t')dt'
\end{equation}
Noting that solution of (\ref{chi-int}) depends on parameter
$q$, let us write it as a power series in $1/q^2$. Introducing
a change of variables $t'\to\tau=t+t'/q$ in the rhs of (\ref{Pc})
we get
$$
\frac{q}{\sqrt{\pi}}\int_{-\infty}^\infty e^{-q^2\tau^2}\chi(t+\tau)d\tau
$$
Now changing variables again $\tau\to\sigma=q\tau$ we get
$$
\frac{1}{\sqrt{\pi}}
  \int_{-\infty}^\infty
  e^{-\sigma^2}
  \chi(t+\frac{\sigma}{q})d\sigma
$$
Finally, expanding $\chi(t+\sigma/q)$ in Taylor series with respect to $\sigma$ we
get
$$
\frac{1}{\sqrt{\pi}}
  \int_{-\infty}^\infty
  e^{-\sigma^2}
  (\chi(t)+\frac{\sigma}{q}\chi'(t)+\frac{\sigma^2}{q^2}\chi''(t)+\cdots)
  d\sigma
$$
All terms which are the odd powers of $\sigma$ vanish after integration.
We obtain
$$
\left(
  -\frac{d^2}{dt^2}+1
\right) \left(
  1+
  \frac{1}{4q^2}\frac{d^2}{dt^2}+
  \cdots
\right)\chi=\chi^3
$$
or
\begin{equation}
\label{chi-exp}
\left(
  -\frac{d^2}{dt^2}+1
\right) e^{\frac{1}{4q^2}~\pd_t^2} \chi=\chi^3,
\end{equation}
where $\exp(1/4q^2~\pd_t^2)$ is understood as a formal expansion
$$
e^{\frac{1}{4q^2}~\pd_t^2}=\sum_{n=0}^{\infty}
  \left(\frac{1}{4q^2}\frac{d^2}{dt^2}\right)^n
$$
Let us note that (\ref{chi-exp}) could be directly obtained from
differential form of the equation, see section \ref{phys}.

In a limit $q\to \infty$ the equation (\ref{chi-exp}) becomes an equation for anharmonic
oscillator. This explains numerically found periodic solutions for large $q$.

\section{Scalar Field Dynamics in Fermionic String Model II}
\label{fermII}

\subsection{System of Nonlinear Integral Equations}

In this section we study the following system of integral equations
\begin{eqnarray}
\label{2f-1}
(\Kc\sigma)(t)&=&\varphi(t)^2\\
\label{2f-2}
(-q^2\pd_t^2+1)(\Kc\varphi)(t)&=&\sigma(t)\varphi(t)
\end{eqnarray}
where the integral operator $\Kc$ is defined by (\ref{kc-def})
and $q$ is a parameter. Physically we are interested in solutions
which are finite on the whole axis and $\varphi(t)$ is odd function
interpolating between $\mp 1$.
This system of equations describes dynamics of the homogenous scalar field
with the lowest mass square (tachyon) in fermionic string model
in the first nontrivial approximation, please see section \ref{phys} for more
physical details.

We are interested in the solutions with the following properties.
The function $\varphi(t)$ is odd and goes to $\mp 1$ on infinities,
i.e. satisfies (\ref{asympt})-(\ref{parity}) and the function $\sigma(t)$
is even and goes to $1$ on infinities, i.e.
\begin{equation}
\sigma(t)=\sigma(-t), ~~\sigma(\mp\infty)=1
\end{equation}

\subsection{Reduction to a Single Equation and Iterative Procedure}

In order to build the solution of the system (\ref{2f-1})-(\ref{2f-2})
first we rewrite it as a single nonlinear integral equation
\begin{equation}
\label{single}
\varphi^2=\Kc\left[ \frac{(-q^2\pd_t^2+1)\Kc \varphi}{\varphi} \right]
\end{equation}
This allows us to construct the following iterative procedure
\begin{equation}
\label{2f-iter}
\varphi_{n+1}(t)=-\varepsilon(t)\sqrt{\Kc
  \left[
    \frac{\Kc_q\varphi_n}{\varphi_n}
  \right]},
\end{equation}
where integral operator $\Kc_q$ is defined by (\ref{integ-q}),
and zero approximation is taken as
\begin{equation}
\label{2f-zero}
\varphi_{0}(t)=-\varepsilon(t),
\end{equation}
where $\varepsilon(t)$ is a step function defined by (\ref{step}).
Please note that although here we discuss iterative procedure for $\varphi(t)$
one can also write a similar procedure for $\sigma(t)$ taking
as a zero approximation a continuous function $\sigma_0(t)=1$.
One could see that (\ref{2f-iter}) assumes that $\varphi_n\neq 0$
that makes us define $\varphi_0$ at $t=0$ in a special way,
see next section for further details.

\subsection{Numerical Results}
\label{num2}

We used iterative procedure (\ref{2f-iter}) to investigate solutions
of the system (\ref{2f-1})-(\ref{2f-2}).

As it was mentioned in the previous section we used $-\varepsilon(t)$
as a zero iteration. Since in (\ref{2f-iter}) we have to inverse
$\varphi_n$ to build the $(n+1)$ iteration we define $\varepsilon(t)$
in the rhs of (\ref{2f-iter}) and (\ref{2f-zero}) to be nonzero for $t=0$.
The basic idea here is to make the value of $\varepsilon(0)$ to
randomly\footnote{The author is grateful to L.V.~Joukovskaya for the idea
to use randomness in deterministic algorithms.}
take values $\pm 1$. In particular we used the following relation
\begin{equation}
\label{eps0}
\varepsilon(0)=(-1)^n
\end{equation}
where $n$ denotes the iteration number. We also tested some more complex
random sequences, but iterative procedure with (\ref{eps0}) was the
fastest and more predictable. One could also take $\varepsilon$ at $t=0$
to be constant for example $\varepsilon(0)=1$ but this lead to some
asymmetry in the resulting solutions.

\begin{figure}
\centering
\includegraphics[width=7cm]{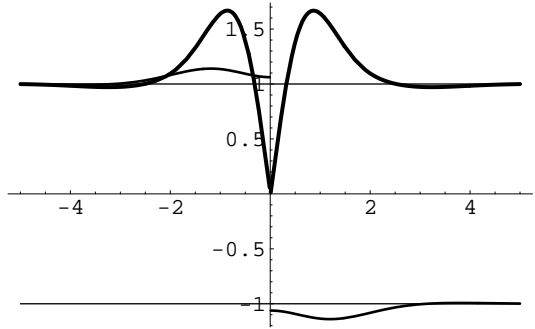}
\caption{Results of iterative procedure solving the system (\ref{2f-1})-(\ref{2f-2}):
$\varphi(t)$ (thin line) and $\sigma(t)$ (thick line) for $q^2=0.96$.}
\label{fig:2f}
\end{figure}

The results of iterative procedure (\ref{2f-iter}) for $q=q_{string}$
are presented on fig.\ref{fig:2f}. Please note that the solution
$\varphi$ has a break at $t=0$.

It was found that there exists a critical value of the parameter
$q^2$: $q^2_{cr}\approx 2.24$. This is the maximum value of $q$
for which there exist interpolating solutions. For $q^2>q^2_{cr}$
the iteration procedure starting from some step faces the negative
argument under the square root in the rhs of (\ref{2f-iter}).
This means that iterative procedure (\ref{2f-iter}) while valid
for finding interpolating solutions fails to find oscillatory ones.

\subsection{Linearization of the System for Large $t$}

In this section we investigate the behavior of the system (\ref{2f-1})-(\ref{2f-2})
for $q\leq q_{cr}$ in the large $t$ limit. The basic idea is analogous to section
\ref{2reg}. We represent the function $\varphi$
which forms the solution for $q\neq 0$ as a deviation from $\varphi_0$ --
the solution for $q=0$. We write a linear integral equation for this deviation
in the large $t$ limit.

Let us write the solution of (\ref{2f-1})-(\ref{2f-2}) as a sum
\begin{equation}
\varphi(t)=\varphi_0(t)+\chi(t),
\end{equation}
where $\varphi_0(t)$ is the solution for $q=0$.
Substituting (\ref{excit}) to (\ref{single}) and leaving
only linear terms in $\chi(t)$ we get
\begin{equation}
\label{chiln2}
\Kc\frac{\Kc_q\varphi_0}{\varphi_0}+\Kc\frac{\Kc_q\chi}{\varphi_0}-
  \Kc\left( \chi\frac{\Kc_q\varphi_0}{\varphi_0^2} \right)=
  \varphi_0^2+2\varphi_0\chi
\end{equation}
Using the fact that $\varphi_0$ solves the system for $q=0$ and
$\varphi_0(t)\simeq 1$ and its derivatives are zero for large $t$
we obtain
\begin{equation}
\label{2f-chi}
\Kc\Kc_q\chi-\Kc\chi=2\chi
\end{equation}
Representing $\chi(t)$ as
$$
\chi(t)=e^{i\Omega t}
$$
and substituting it to (\ref{2f-chi}) we get the following
characteristic equation
\begin{equation}
\label{2f-oq}
(q^2\Omega^2+1)e^{-\frac{1}{2}\Omega^2}-e^{-\frac{1}{4}\Omega^2}=2
\end{equation}
Analogously to what we did in section \ref{2reg} we consider
(\ref{2f-oq}) as an equation for a complex variable
$\Omega$ with parameter $q^2$.
There is a minimum value of $q^2$: $q^2_0\approx 3.05$ for which
$\Omega$ is real.
For $q^2<q^2_0$ it has solutions
with nonzero imaginary parts which gives oscillatory regime with
exponentially decreasing amplitude. For $q^2>q^2_0\approx3.05$ solutions are
real. As it was mentioned in the previous section numerical
computations give the critical value $q^2_{cr}\approx 2.24$.
This difference probably means that one can not put $\varphi_0$
equal to $1$ in the rhs of (\ref{2f-chi}) or even consider nonlinear
terms in $\chi$ in (\ref{chiln2}).

\subsection{Asymptotic Behavior For Large $q$}

Here we investigate asymptotic behavior of the system (\ref{2f-1})-(\ref{2f-2})
for large $q$. As in the section \ref{big-q} we introduce a change of variables
analogous to (\ref{chi-Psi})
$$
\chi(t)=\varphi(qt)
$$
$$
\xi(t)=\sigma(qt)
$$
In terms of $\chi(t)$ and $\xi(t)$ the system rewrites
\begin{eqnarray}
\label{cx-1}
(\Pc_q\xi)(t)&=&\chi(t)^2\\
\label{cx-2}
(-\pd_t^2+1)(\Pc_q\chi)(t)&=&\xi(t)\chi(t),
\end{eqnarray}
where integral operator $\Pc_q$ is defined by (\ref{Pc}).

Performing computations analogous to section \ref{big-q} we get that in the large
$q$ limit the system (\ref{cx-1})-(\ref{cx-2}) becomes
\begin{eqnarray}
\xi(t)&=&\chi(t)^2\\
(-\pd_t^2+1)\chi(t)&=&\xi(t)\chi(t)
\end{eqnarray}
that is equivalent to an anharmonic oscillator.

\newpage
\section{Differential Form of the Equations and Physical Roots}
\label{phys}

In this section we provide physical details including
actions for which corresponding equations of motion form the
subject of this paper.

Effective p-adic action is given by \cite{BFOW,FramPadic,VVZ}
\begin{equation}
\label{p-action}
S=\frac{1}{g_p^2}\int d^dx
\left[
  -\frac{1}{2}\phi p^{-\frac{1}{2}\Box} \phi + \frac{1}{p+1}\phi^{p+1}
\right],~~\frac{1}{g_p^2}\equiv\frac{1}{g^2}\frac{p^2}{p-1}
\end{equation}
Here $\phi$ is a scalar field which describes tachyon in p-adic string model,
$x=(t,\overrightarrow{x})$ are $d$-dimensional space-time coordinates,
$p$ is a prime number, $g_p$ is a coupling constant ($g$ is universal coupling
constant), the D'Alamber operator is defined in a standard way
\begin{equation}
\Box=-\frac{\partial^2}{\partial t^2} + \nabla\cdot\nabla,
\end{equation}
and operator $p^{-\frac{1}{2}\Box}$ is understood in the sense of expansion
\begin{equation}
p^{-\frac{1}{2}\Box}=e^{-\frac{1}{2}\ln p~\Box}=
  \sum_{n=0}^\infty \left( -\frac{1}{2}\ln p \right)^n \frac{1}{n!} \Box^n
\end{equation}
The corresponding equations of motion are the following
\begin{equation}
\label{p-eom}
p^{-\frac{1}{2}\Box}\phi=\phi^p
\end{equation}
For homogenous field configurations (\ref{p-eom}) rewrites as follows
\begin{equation}
\label{padic-eom}
p^{\frac{1}{2}\partial^2_t}\phi=\phi^p
\end{equation}

If we consider slowly varying solutions and neglect high
order derivatives in the lhs of (\ref{padic-eom})
we get equations for an anharmonic oscillator
\begin{equation}
\label{padic-osc}
\frac{1}{2}\ln p~\partial^2_t \phi+\phi=\phi^p,
\end{equation}
in the potential
\begin{equation}
\label{padic-pot}
V(\phi)=\frac{2}{\ln p}
  \left(
    \frac{1}{2}\phi^2-\frac{1}{p+1}\phi^{p+1}
  \right)
\end{equation}
From (\ref{padic-pot}) we see that the cases with even and odd
$p$ we have qualitatively different behaviors \cite{MolZw}.

\begin{figure}[!ht]
\centering
\includegraphics[width=5cm]{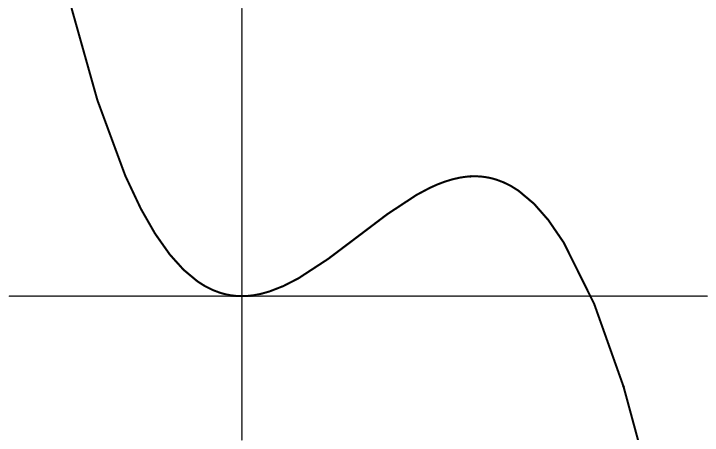}
\hspace{1cm}
\includegraphics[width=5cm]{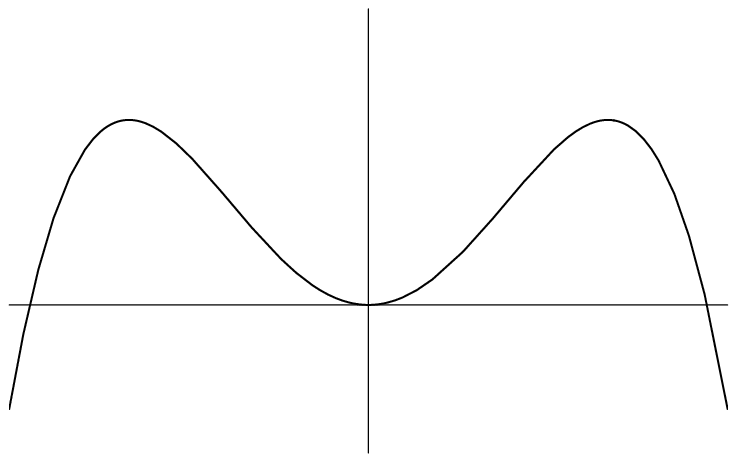}
\caption{p-adic potential for the case $p=2$ (left)
and $p=3$ (right).}
\label{fig:pot-23}
\end{figure}
In particular in the case $p=2$ the potential (\ref{padic-pot})
has minimum when $\phi=0$ (fig.\ref{fig:pot-23}) and maximum
when $\phi=1$. In \cite{MolZw} it was proved that there are
no monotonic solutions interpolating between these points.

In the case $p=3$ the potential has three extremal points
\begin{equation}
\label{eqk-even}
\phi_0^{(1)}=1,~~~~\phi_0^{(2)}=0,~~~~\phi_0^{(3)}=-1
\end{equation}
The extremal points $\phi_0^{(1,3)}$ correspond to unstable
vacua, and $\phi_0^{(2)}$ corresponds to the stable one.
In the previous sections we were interested in the time dependant
solutions symmetrically interpolating between $\phi_0^{(1)}$ and
$\phi_0^{(3)}$.

The action for the scalar field with lowest mass square (tachyon)
in fermionic string model in the first nontrivial approximation
is given by \cite{AJK}
\begin{equation}
\label{tacha}
S[\upsilon,\psi]=
  \int\;dt
  \left[
    \frac14\upsilon(t)^2+
    \frac{q^2}{2}(\pd_t\psi(t))^2+
    \frac12\psi^2(t)-
    \frac12\Upsilon(t)\Psi^2(t)
  \right],
\end{equation}
where
\begin{equation}
\label{UP}
\Upsilon(t)=\exp(-\frac18\pd^2)\upsilon(t),~~~~~
\Psi(t)=\exp(-\frac18\pd^2)\psi(t)
\end{equation}
and
$$
q^2=q^2_{string}=-\frac{1}{4\ln\gamma}\approx 0.96
$$
Here $\psi$ is the tachyon filed and $\upsilon$ is the auxiliary
filed. We will write the equations of motion in terms of $\Upsilon$
and $\Psi$ while the physical fields $\psi$ and $\upsilon$
will be obtained using the inverse of (\ref{UP}).

First let us consider an approximation for the action (\ref{tacha})
which leads to simpler equations of motion.
If we assume that we can neglect the smoothness of auxiliary
field, i.e. in the interacting term write $\upsilon$
instead of $\Upsilon$ we get the following approximate action \cite{AJK}
\begin{equation}
\label{tachapp}
S[\upsilon,\psi]=
  \int\;dt
  \left[
    \frac14\upsilon(t)^2+
    \frac{q^2}{2}(\pd_t\psi(t))^2+
    \frac12\psi^2(t)-
    \frac12\upsilon(t)\Psi^2(t)
  \right]
\end{equation}
This action leads to the following equations of motion
\begin{eqnarray}
\upsilon(t)&=&\Psi^2(t)\\
(-q^2\pd_ò^2+1)e^{\frac14\pd^2_t}\Psi(t)&=&\upsilon(t)\Psi(t)
\end{eqnarray}
The first equation gives the expression for $\upsilon$ in terms of $\Psi$,
substituting it to the second one we get
\begin{equation}
\label{tach-appr}
(-q^2\pd_ò^2+1)e^{\frac14\pd^2_t}\Psi(t)=\Psi(t)^3
\end{equation}
If we rewrite (\ref{tach-appr}) in the integral form
(see (\ref{intdiff}) with $a=1/4$) we get
\begin{equation}
(-q^2\pd_ò^2+1)\Kc\Psi=\Psi^3
\end{equation}
This equation was studied in section \ref{fermI}, see equation
(\ref{ferm}).

The equations of motion for original action (\ref{tacha})
are the following
\begin{eqnarray}
\label{tach-1}
e^{\frac14\pd^2}\Upsilon(t)&=&\Psi^2(t)\\
\label{tach-2}
(-q^2\pd_ò^2+1)e^{\frac14\pd^2}\Psi(t)&=&\Upsilon(t)\Psi(t),
\end{eqnarray}
In the integral form this system is equivalent to the system
(\ref{2f-1})-(\ref{2f-2}) which was studied in section \ref{fermII}.

\begin{figure}[!ht]
\centering
\includegraphics[width=7cm]{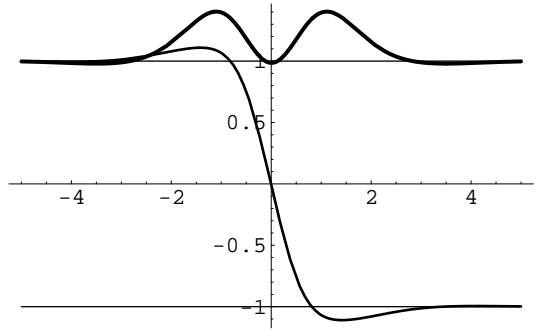}
\caption{Physical fields $\psi(t)$ (thin line) and $\upsilon(t)$ (thick line) are
smooth for all $t$ ($q^2=q^2_{string}$).}
\label{fig:cont}
\end{figure}

Let us compare the dynamics of physical field $\psi(t)$ obtained from
exact action (\ref{tacha}) and approximate action (\ref{tachapp}).
We compute physical field $\psi(t)$ from solutions of (\ref{tach-appr})
and (\ref{tach-1})-(\ref{tach-2}) using the inverse of (\ref{UP})
which has the following form
\begin{equation}
\label{smooth}
\psi(t)=e^{\frac18\pd^2}\Psi(t)=\sqrt{\frac{2}{\pi}}\int_{-\infty}^{\infty}
  e^{-2(t-t')^{2}} \Psi(t')dt'
\end{equation}
Note that although $\Psi$ solving (\ref{tach-1})-(\ref{tach-2})
has a break at $t=0$ (see section \ref{num2}) the application
of (\ref{smooth}) makes the resulting physical filed $\psi$
smooth, see fig.\ref{fig:cont}.

\begin{figure}[!ht]
\centering
\includegraphics[width=7cm]{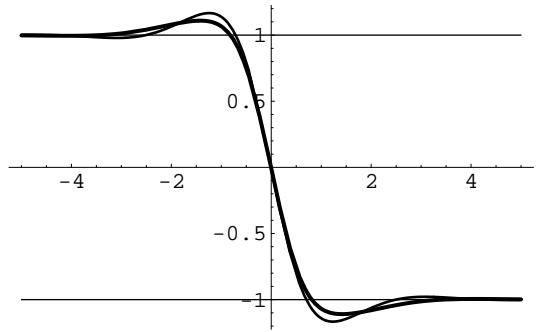}
\caption{Comparison of physical field dynamics $\psi(t)$ obtained from
exact action (\ref{tacha}), thick line, and approximate action
(\ref{tachapp}), thin line, for $q^2=q^2_{string}$.}
\label{fig:cmp}
\end{figure}

The results of comparison of physical fields obtained from exact
and approximate actions are presented on fig.\ref{fig:cmp}.

\noindent{\bf\Large Conclusions}

\noindent In this paper we numerically studied nonlinear equations
depending on one variable with infinite number of derivatives.
Two different regimes for solution of equations of motion for
the fermionic string model are found and the critical value
of the parameter is derived. It would be interesting to
explore in more detail the transition between two regimes.
Also generalization of the results of the paper to the case of
several variables would be important.

\newpage
\noindent{\bf\Large Acknowledgments}

\noindent This work was partly supported by RFFI-02-01-01084.
The author is grateful to
I.Ya.~Aref'eva, V.V.~Belokurov, D.P.~Demichev, B.~Dragovich,
L.V.~Jou\-kov\-ska\-ya, A.Yu.~Khrennikov and I.V.~Volovich for fruitful discussions.


\begin{thebibliography}{99}

\bibitem{Vla} V.S. Vladimirov, {\it Equations of Mathematical Physics}, Nauka,
Moskow, 1988

\bibitem{BFOW} L.~Brekke, P.G.O.~Freund, M.~Olson, E.~Witten,
{\it Non-archimedian string dynamics}, Nucl.Phys. {\bf B302}
(1988) 365.
%%CITATION = NUPHA,B302,365;%%

\bibitem{FramPadic} P.H. Frampton, Y.Okada, {\it Effective scalar
field theory of $p$-adic string}, Phys.Rev. {\bf D37} (1989)

\bibitem{VVZ} V.S.~Vladimirov, I.V.~Volovich and E.I.~Zelenov,
{\it $P$-adic Analysis and  Mathematical Physics}, World Sci.
1994.

\bibitem{Kh} A.Yu.~Khrennikov, {\it $p$-adic valued distributions in
mathematical physics.} Kluwer Acad. Publ., Dordrecht, 1994.

\bibitem{MolZw}
N.~Moeller and B.~Zwiebach, {\it Dynamics with infinitely many
time derivatives and rolling tachyons}, hep-th/0207107.\\
%%CITATION = HEP-TH 0207107;%%
H.Yang, {\it Stress tensors in p-adic string theory and truncated OSFT},
hep-th/0209197.
%%CITATION = HEP-TH 0209197;%%

\bibitem{Sen} D.~Ghoshal, A.~Sen, {\it Tachyon Condensation and Brane
Descent Relations in p-adic String Theory},
Nucl.Phys. {\bf B}584 (2000), pp.300-312, hep-th/0003278
%%CITATION = HEP-TH 0003278;%%

\bibitem{AJK} I.Ya.~Aref'eva, L.V.~Joukovskaya and A.~S.~Koshelev,
{\it Time Evolution in Superstring Field Theory on
non-BPS brane. I. Rolling Tachyon and Energy-Momentum Conservation},
hep-th/0301137.
%%CITATION = HEP-TH 0301137;%%

\bibitem{Gamma} E. Gamma, R.Helm, R.Johnson and J. Vlissides, {\it Design Patterns. Elements of
Reusable Object-Oriented Software}, Addison-Wesley, 1995

\bibitem{Eliens} A. Eli\"{e}ns,  {\it Principles of Object-Oriented Software Development},
 Addison-Wesley, 2000

\bibitem{Kr} {\it Functional Analysis}, Ed. S.G.~Krein, Nauka, Moscow, 1972

\end{thebibliography}
\end{document}